\RequirePackage{fix-cm}
\documentclass[smallextended]{svjour3}       
\smartqed  
\usepackage{graphicx}

\usepackage{graphicx}
\usepackage{amssymb}
\usepackage{amsmath}
\usepackage{cite}
\usepackage{epstopdf}
\usepackage{cite}
\usepackage{hyperref}
\usepackage{balance}
\usepackage{moreverb}
\usepackage{xcolor}
\usepackage[dvips]{epsfig}
\usepackage[latin1]{inputenc}
\usepackage[T1]{fontenc}
\usepackage{subfigure,amsfonts,balance}
\usepackage{amsmath}
\usepackage{bbm}
\usepackage{bm,cite,algorithm,algorithmic,float,amsmath,amssymb}
\usepackage{color}
\usepackage[dvips]{graphicx}

\usepackage{bm,cite,algorithm,algorithmic,float,amsmath,amssymb}

\makeatletter
\def\ScaleIfNeeded{%
\ifdim\Gin@nat@width>\linewidth \linewidth \else \Gin@nat@width
\fi } \makeatother

\DeclareMathOperator{\tr}{\mathrm{tr}}
\DeclareMathOperator{\rank}{\mathrm{rank}}
\DeclareMathOperator{\E}{\mathbb{E}}

\begin{document}

\title{Particle Swarm Optimization for Weighted Sum Rate Maximization in MIMO Broadcast Channels}


\author{Tung Thanh Vu \and Ha Hoang Kha \and Trung Quang Duong \and Nguyen-Son Vo}


\institute{Tung Thanh Vu and Nguyen-Son Vo \at
                Duy Tan University, Da Nang, Vietnam \\
                \email{\{vuthanhtung1, vonguyenson\}}@dtu.edu.vn          
           \and
           Tung Thanh Vu and Ha Hoang Kha \at
                Ho Chi Minh City University of Technology, Vietnam \\
                \email{hhkha@hcmut.edu.vn}
           \and
           Trung Quang Duong \at
                Queen's University Belfast\\
			    \email{trung.q.duong@qub.ac.uk}.
			}
\date{Received: date / Accepted: date}

\maketitle

\begin{abstract}
In this paper, we investigate the downlink multiple-input-multiple-output (MIMO) broadcast channels in which a base transceiver station (BTS) broadcasts multiple data streams to $K$ MIMO mobile stations (MSs) simultaneously. In order to maximize the weighted sum-rate (WSR) of the system subject to the transmitted power constraint, the design problem is to find the pre-coding matrices at BTS and the decoding matrices at MSs. However, such a design problem is typically a nonlinear and nonconvex optimization and, thus, it is quite hard to obtain the analytical solutions. To tackle with the mathematical difficulties, we propose an efficient stochastic optimization algorithm to optimize the transceiver matrices. Specifically, we utilize the linear minimum mean square error (MMSE) Wiener filters at MSs. Then, we introduce the constrained particle swarm optimization (PSO) algorithm to jointly optimize the precoding and decoding matrices. Numerical experiments are exhibited to validate the effectiveness of the proposed algorithm in terms of convergence, computational complexity and total WSR.

\end{abstract}

\section{Introduction}
\label{sec:introd}
Multiple-input multiple-output (MIMO) transmission has been adopted for a various wireless communications standards such as the long-term evolution (LTE) and WiMax since it enables the enhancement of communication reliability and channel capacity in wireless networks without the requirement of additional bandwidth or power \cite{Tela99, Fosc98, Gold03}. The single user MIMO communication channels have extensively investigated in the literature, see, e.g. \cite{Scut09} and references therein. Recently, the multiuser MIMO channels have drawn much research attention for efficiently utilizing the spectrum and improving the system capacity \cite{Gesb07}. In this paper, we study an important scenario in which a base transceiver station (BTS) equipped with multiple antennas sends different information streams to the multiple MSs, each equipped with multiple antennas. Such a system model is also well-known as downlink multi-user MIMO broadcast channels (BCs), in which the major challenge is how to suppress co-channel interference at MSs.

To mitigate interference and enhance the transmission data rate in downlink multiuser MIMO wireless networks, several linear precoding strategies using different algorithms have been studied \cite{Wein06, Shen06, Spen04, Zhou08, Bin10, Bin10-2, Gan06, Yu04}. It was shown in \cite{Wein06,Shen06} that dirty paper coding (DPC) can achieve the capacity region of the multiuser MIMO BCs. Although DPC is the optimal transmission scheme for MIMO BCs, it is high complex for practical implementation. The authors in \cite{Shen06, Spen04} have also proved that block diagonalization (BD) schemes can achieve the suboptimal transmission with lower complexity. Basically, BD is a version of the zero-forcing method in which the precoders are designed to null inter-user interference. Hence, BD provides both high per-MS rates and high sum rates by combining the benefits of spatial multiplexing and multiuser MIMO systems. However, to efficiently suppress interference, the BD schemes require the number of transmit antennas at BTS to be larger than the sum of receive antennas of all MSs \cite{Chen07}. This makes BD inapplicable to the systems where there are a large number of MIMO MSs or the deployment of the large number of transmit antennas at BTS is not applicable. Alternative approaches are based on deterministic optimization algorithms to maximize the system channel capacity \cite{Gan06, Yu04}. Due to the nonconvex nature of the design problem, the globally optimal solutions are generally not guaranteed. In addition, the deterministic iterative algorithms highly rely on the step size and/or gradient computation. As opposed to the previously existing methods, in this paper, we directly solve the weighted sum-rate (WSR) maximization problem by employing the stochastic global optimization approach namely particle swarm optimization (PSO).

The PSO algorithm is known as a class of swarm intelligence methods for solving global optimization problems. Introduced by Kennedy and Eberhart in 1995 \cite{Kenn95}, the key idea of PSO was inspired by social behaviour of bird flocks and fish schools. PSO seeks the optimal solution by transferring information between particles and retains the global search strategy based on swarms. PSO can be efficiently implemented  and take a reasonably short time to converge to the optimal solution \cite{Chen10}. Hence, PSO has been exploited in many technical applications \cite{Poli07,  Chen10, Fang08}. Reference \cite{Fang08} applied PSO to find optimal beamforming for multiuser MIMO systems. The algorithm in \cite{Fang08} aims to improve the total signal-to-interference-plus-noise (SINR) at each receiver and restricts single data steam to be transmitted to each MS. Motivated by the previous works, we develop the PSO algorithm to search the optimal solution for the nonconvex and intractable optimization problems of linear transceiver designs in downlink multiuser MIMO interference channels. In particular, we first formulate the design problems of finding the linear precoding and decoding matrices which maximize the total WSR of the systems subject to the total power constraints. From the conventional PSO algorithm which is normally applied to unconstrained optimization problem, we developed the constrained PSO algorithm which keeps the search space on the feasible regions by projecting the particles into feasible sets. The constrained PSO algorithm is used to find the percoding matrices at BTS while the decoding matrices at MSs are the linear minimum mean square error (MMSE) Wiener filter solutions. As compared to the BD method which imposes the constraint on the number of antennas, the proposed method can be applied for more general scenarios of downlink MIMO systems. The computational complexity of the proposed PSO algorithm is shown to be in polynomial-time. The performance of the proposed algorithm shall be validated by numerical simulation results in compared with the BD method which is proven to achieve the suboptimal transmission \cite{Shen06, Spen04}.

The rest of this paper is organized as follows.  Section \ref{sec:Model} introduces the downlink multiuser MIMO interference channels and presents the use of BD method for the design of the precoders and decoding matrices. In Section \ref{sec:ProposedMethod}, we present the PSO algorithm to find the optimal transceivers. The numerical results are provided in Section \ref{sec:Results}. Finally, Section \ref{sec:Conclusion} presents concluding remarks.

\emph{Notation}: Matrices and vectors are respectively denoted by boldface capital and lowercase letters. The transposition and conjugate transposition of complex matrix $\pmb{X}$ are $\pmb{X}^T$ and $\pmb{X}^H$, respectively. The $i$-th column of matrix $\pmb{X}$ is denoted by $\pmb{X}(:,i)$. $\pmb{I}$ and $\pmb{0}$ are respectively identity and zero matrices with the appropriate dimensions. $\tr (.)$, $\rank (.)$ and $\E(.)$  indicate the trace, rank and expectation operators, respectively. $||\pmb{x}||_2$ denotes the Euclidean norm while $||\pmb{X}||_F$ presents the Frobenius norm.
$\pmb{x}\sim \mathcal{CN}(\bar{\pmb{x}},\pmb{R}_{\pmb{x}})$ mean that $\pmb{x}$ is a complex Gaussian random vector with means $\bar{\pmb{x}}$ and covariance $\pmb{R}_{\pmb{x}}$.
\section{System Model and BD Method}
\label{sec:Model}
In this section, we first present the model of downlink MIMO interference channels. Then, we briefly introduce the BD scheme to handle interference and maximize WSR.
\subsection{System model for downlink MIMO interference channels}
 \begin{figure}[htb!]
\begin{center}
\epsfxsize=10cm
\leavevmode\epsfbox{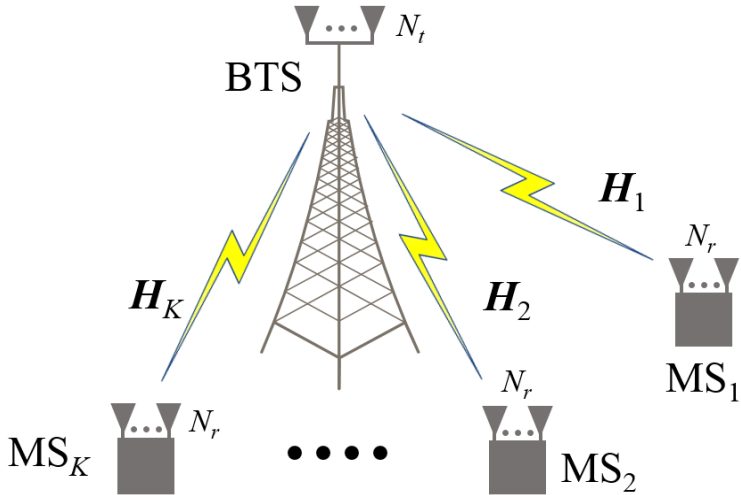}
\caption{A system model of downlink MIMO interference channels.}
\label{fig:PSO42}
\end{center}
\end{figure}
A downlink MIMO interference wireless system is considered as in Fig. \ref{fig:DownlinkMIMOIC}, in which one BTS equipped with $N_t$ antennas sends signals to $K$ mobile stations (MSs), each equipped with the $N_{r_k}$ antennas, for $k=1,...,K$. For the sake of presentation and without loss of generality, we assume that all MSs are equipped with the same number of antennas $N_{r_k}=N_r$ for all $k$. Each MS receives $d_k=d$ data streams. Let $\pmb{x}_k \in \mathbb{C}^{d \times 1}$ be the signal vector transmitted from BTS to the $k$-th MS. Before the signal is transmitted over $N_t$ antennas, a linear precoding matrix $\pmb{F}_k \in \mathbb{C}^{N_t \times d}$ is applied on $\pmb{x}_k$ and then takes sum for all MSs to form a signal vector $\pmb{s} \in \mathbb{C}^{N_t \times 1}$, given by $\pmb{s} = \sum\limits_{k=1}^K\pmb{F}_k\pmb{x}_k$. Hence, the transmitted power at BTS can be defined by
\begin{align}\label{P:k:ori}
P = \E\{ \left\| \pmb{s} \right\|^2 \} = \sum\limits_{k=1}^K\text{tr}\left( {{\pmb{F}_k}\E\{ {{\pmb{x}_k}\pmb{x}_k^H} \}\pmb{F}_k^H} \right).
\end{align}
Without loss of generality, we assume that $\E\{ {{\pmb{x}_k}\pmb{x}_k^H} \} = {\pmb{I}_{{d}}}$ and, hence, the constraint on the transmitted power from Eq. \eqref{P:k:ori} can be rewritten as
\begin{align}\label{P:k}
P = \sum\limits_{k=1}^K\text{tr}\left( {{\pmb{F}_k}\pmb{F}_k^H} \right) \leq P_{max}
\end{align}
where $P_{max}$ is the maximum allowable transmitted power at the BTS.

In this paper, we consider the static flat-fading MIMO channels where the channel matrix from the BTS to the $k$-th MS is denoted by $\pmb{H}_k \in \mathbb{C}^{N_r \times N_t}$. We assume that all users are co-located and, hence, large scale fading can be excluded \cite{Riha14}. The channel entries are independent and identical distributed (i.i.d.) complex Gaussian variables and their magnitudes follow the Rayleigh distribution. The channel is assumed to be a block-fading which remains unchanged for a frame duration and varies independently for every frame. Since all signals are broadcasted from BTS, each MS receives  not only its desired signal but also unintended signals from the other MSs. The received signal at the $k$-th MS can be expressed as
\begin{align}\label{Signal:Rx}
\pmb{y}_k = \pmb{H}_k\pmb{s} = \underbrace{\pmb{H}_k\pmb{F}_k\pmb{x}_k}_{\text{desired signal}} + \underbrace{\sum\limits_{\ell = 1,\ell \ne k}^K \pmb{H}_k\pmb{F}_\ell\pmb{x}_\ell}_{\text{inter-user interference}}  + \underbrace{\pmb{n}_k}_{\text{noise}}
\end{align}
where $\pmb{y}_k \in \mathbb{C}^{N_r \times 1}$ is the received signal at the $k$-th MS and $\pmb{n}_k \sim \mathcal{CN}\left( 0,\sigma _k^2\pmb{I}_{N_r} \right)$ is a complex Gaussian noise vector at the $k$-th MS.
To recover the desired signal, the MSs apply the decoding matrix $\pmb{W}_k \in \mathbb{C}^{N_r \times d}$ to the received signal yielding
\begin{align}\label{RecoveredSignal}
\widehat {\pmb{x}}_k &= \pmb{W}_k^H\pmb{y}_k \\ \nonumber
& = \underbrace{\pmb{W}_k^H\pmb{H}_k\pmb{F}_k\pmb{x}_k}_{\text{desired signal}} +  \underbrace{\sum_{\ell = 1,\ell \ne k}^K \pmb{W}_k^H\pmb{H}_k\pmb{F}_\ell\pmb{x}_\ell}_{\text{inter-user interference}}  + \underbrace{\pmb{W}_k^H\pmb{n}_k}_{\text{noise}}.
\end{align}
Channel capacity at the $k$-th MS can be calculated as \cite{Bazz12}
\begin{align}\label{Cap:MSk}
\mathcal{R}_k = \log _2\left| \pmb{I}_d + \pmb{W}_k^H\pmb{H}_k\pmb{F}_k\pmb{F}_k^H\pmb{H}_k^H\pmb{W}_k\pmb{R}_{z_k}^{- 1} \right|
\end{align}
where $\pmb{R}_{z_k} = \pmb{W}_k^H\left(\sum\limits_{\ell = 1,\ell \ne k}^K \pmb{H}_k{\pmb{F}_\ell\pmb{F}_\ell^H\pmb{H}_k^H}  + \sigma _k^2\pmb{I}_{N_r}\right)\pmb{W}_k$ is the correlation matrix of interference and noise in Eq.\eqref{RecoveredSignal}. The total WSR over $K$ MSs can be expressed as
\begin{align}\label{Cap:total}
{\mathcal{R}} = \sum\limits_{k = 1}^K {\omega_k{\mathcal{R}}_k}
\end{align}
where $\omega_k \geq 0$ is the weight factor to present the different priority for the MSs.  The problem of interest is to design the precoding matrices $\pmb{F}_k$ and post-processing matrices $\pmb{W}_k$ in order to obtain the maximum WSR over $K$ MSs.

\subsection{Block diagonalization for downlink multiuser MIMO interference channels}
An efficient method to cancel the interference in MIMO downlink BCs is block diagonalization \cite{Spen04, Shen06}. This subsection introduces the block diagonalization method for maximizing WSR which will be used as a benchmark for performance comparison. The key idea of the BD strategy is to find the precoding matrices $\pmb{F}_k$ and decoding matrices $\pmb{W}_k$ to force all inter-user interference to zero and maximize the total channel capacity. From Eq. \eqref{RecoveredSignal}, the interference rejection condition must be expressed as
\begin{align} \label{IAcondition}
\pmb{H}_k\pmb{F}_\ell = \pmb{0},\,\,\,\,\,\, \forall k, \text{and}\,\, \forall {\ell \neq k}.
\end{align}
Let us define the interference matrix as follows
\begin{align}\label{InterferenceMatrix}
\pmb{\mathcal{H}}_k = \left[ \pmb{H}_1^T,\pmb{H}_2^T,\cdots,\pmb{H}_{k-1}^T,\pmb{H}_{k+1}^T,\cdots,\pmb{H}_K^T  \right]^T
\end{align}
where $\pmb{\mathcal{H}}_k \in \mathbb{C}^{(K-1)N_r \times N_t}$. Then, Eq. \eqref{IAcondition} can be equivalently rewritten as
\begin{align} \label{IAcondition2}
\pmb{\mathcal{H}}_k\pmb{F}_k = \pmb{0},\,\,\,\,\,\, \forall k.
\end{align}
It means that $\pmb{F}_k$ has to lie in the null space of $\pmb{\mathcal{H}}_k$. For the existence of $\pmb{F}_k$ satisfying Eq. \eqref{IAcondition2}, we must have $N_t-(K-1)N_r \geq d$. The singular value decomposition (SVD) of $\pmb{\mathcal{H}}_k$ can be expressed as
\begin{align} \label{intH:SVD}
\pmb{\mathcal{H}}_k = \pmb{U}_k\pmb{\Sigma}_k\pmb{V}_k^H
\end{align}
where $\pmb{\Sigma}_k$ is the diagonal matrix with diagonal elements being singular values in decreasing order. Define $\pmb{B}_k \in \mathbb{C}^{N_t \times d}$ as the matrix of the last $d$ columns in matrix $\pmb{V}_k$, i.e., $\pmb{B}_k$ lies on the null space of $\pmb{\mathcal{H}}_k$. Thus, the precoder with $\pmb{B}_k$ can make each block $\pmb{H}_k\pmb{B}_k$ of the k-th MS free from inter-user interference. In order to separate the block channel $\pmb{H}_k\pmb{B}_k$ into $N_r$ parallel sub-channels, an SVD is applied to $\pmb{H}_k\pmb{B}_k$ as follows
\begin{align} \label{HB:SVD}
\pmb{H}_k\pmb{B}_k = \pmb{\tilde{U}}_k\pmb{\Lambda}_k\pmb{\tilde{V}}_k^H
\end{align}
where $\pmb{\Lambda}_k$ is the diagonal matrix of decreasing singular values of $\pmb{H}_k\pmb{B}_k$. Define $\pmb{D}_k \in \mathbb{C}^{d \times d}$ as a matrix of the first $d$ columns of $\pmb{\tilde{V}}_k$. Now, the precoder is defined as
\begin{align}
\pmb{F}_k=\pmb{B}_k\pmb{D}_k\pmb{P}_k^\frac{1}{2},
\end{align}
where $\pmb{P}_k=\mathrm{diag}[p_{k,1},p_{k,2},\ldots, p_{k,d}]$ is the power allocation matrix. The decoding matrix $\pmb{W}_k$ is chosen as the first $d$ columns of $\pmb{\tilde{U}}_k$. Eq. \eqref{RecoveredSignal} becomes
\begin{align} \label{RecoverdSignal:BD}
\widehat{\pmb{x}}_k = \pmb{W}_k^H\pmb{H}_k\pmb{B}_k\pmb{D}_k\pmb{P}_k^\frac{1}{2}\pmb{x}_k + \pmb{W}_k^H\pmb{n}_k.
\end{align}
Then, channel capacity at the $k$-th MS using BD can be expressed as
\begin{align}
\mathcal{R}_{k,BD} = \sum\limits_{i=1}^{d} \log _2 (1 + \frac{\pmb{\Lambda}_k^2(i,i)}{\sigma_k^2}p_{k,i}).
\end{align}
The total WSR over $K$ MSs can be expressed as
\begin{align}\label{SumRate:BD}
\mathcal{R} = \sum\limits_{k = 1}^K \omega_k \mathcal{R}_{k,BD} = \sum\limits_{k = 1}^K\sum\limits_{i=1}^{d} \omega_k \log _2 (1 + \frac{\pmb{\Lambda}_k^2(i,i)}{\sigma_k^2}p_{k,i})
\end{align}
 In order to maximize the WSR in Eq. \eqref{SumRate:BD}, we need to find the optimal $\pmb{P}_k$ by solving the following optimization problem
\begin{subequations}\label{PkOpt}
\begin{eqnarray}
& \underset{p_{k,i}}{\text{max}} & \sum\limits_{k = 1}^K \sum\limits_{i=1}^{d}\omega_k \log _2 (1 + \frac{\pmb{\Lambda}_k^2(i,i)}{\sigma_k^2}p_{k,i}) \label{obPkOpt} \\
& \text{s.t.} & \sum\limits_{k = 1}^K\sum\limits_{i = 1}^d p_{k,i} \leq P_{\max}.
\end{eqnarray}
\end{subequations}
The Lagrangian of problem \eqref{PkOpt} can be obtained as
\begin{align}
\mathcal{L}(p_{k,i},\lambda)  = & - \sum\limits_{k = 1}^K \sum\limits_{i=1}^{d} \omega_k \log _2 (1 + \frac{\pmb{\Lambda}_k^2(i,i)}{\sigma_k^2}p_{k,i})\\ \nonumber & + \lambda (\sum\limits_{k = 1}^K\sum\limits_{i = 1}^d p_{k,i} - P_{max})
\end{align}
where $\lambda$ is the Lagrangian dual variables associated with the power constraint at BTS. Since problem \eqref{PkOpt} is obviously convex optimization, the optimal solutions of problem \eqref{PkOpt} must satisfy the set of Karush-Kuhn-Tucker (KKT)
\begin{subequations}\label{KKT}
\begin{eqnarray}
\frac{-\omega_k\frac{\pmb{\Lambda}_k^2(i,i)}{\sigma_k^2}}{(1+\frac{\pmb{\Lambda}_k^2(i,i)}{\sigma_k^2}p_{k,i})\ln{2}}+\lambda = 0 \\
\sum\limits_{k = 1}^K\sum\limits_{i = 1}^d p_{k,i} = P_{max}\\
\lambda \geq 0.
\end{eqnarray}
\end{subequations}
Then, the solution is given by
\begin{equation}\label{proresult}
p_{k,i}=\max\bigg(0,\frac{\omega_k}{\lambda\ln{2}}-\frac{\sigma_k^2}{\pmb{\Lambda_k}^2(i,i)}\bigg)
\end{equation}
Finally, the water-filling (WF) algorithm in \cite{Scut09} is applied to find optimal $\lambda$ under the total power constraint at BTS.
\section{PSO for Downlink Multiuser MIMO Interference Channels}
\label{sec:ProposedMethod}
The purpose of the current paper is to directly maximize the WSR of the networks in \eqref{Cap:total} by searching the optimal precoding and decoding matrices. In order to reduce the search space and complexity, we adopt the MMSE Wiener filter at MSs since they take the effects of noise into account and are typically exploited at the receivers because of their simplicity \cite{Bazz12}. Using Eq. \eqref{RecoveredSignal}, the decoding matrix $\pmb{W}_k$ can be found by \cite{Kay93} as
\begin{align}\label{Wk:Wiener}
\pmb{W}_k = {\left(\sum\limits_{\ell = 1}^K {\pmb{H}_k{\pmb{F}_\ell}\pmb{F}_\ell^H\pmb{H}_k^H}  + \sigma _k^2{\pmb{I}_{{N_{{r_k}}}}}\right)^{ - 1}}{\pmb{H}_k}{\pmb{F}_k}.
\end{align}
Then, we find the precoding matrices $\pmb{F}_k$ by solving the optimization problem as follows
\begin{subequations}\label{FkOpt}
\begin{eqnarray}
& \underset{{\left\{ {{\pmb{F}_k}} \right\}_{k = 1}^K}}{\text{max}} & \sum\limits_{k = 1}^K {\omega_k{\mathcal{R}}_k} \label{objFkOpt}\\
& \text{s.t.} & \sum\limits_{k=1}^K\tr(\pmb{F}_k \pmb{F}_k^H) \leq P_{max}.
\end{eqnarray}
\end{subequations}
The problem in Eq. \eqref{FkOpt} is the NP-hard nonconvex optimization which renders the mathematical challenges in  finding the optimal solutions by the deterministic optimization. This motivates us to develop the PSO algorithm to search for the global solutions.

The standard PSO is originally known as a stochastic global optimization for unconstrained optimization problems. The PSO algorithm follows the simulation of social particle behaviors in which particles are coordinated to share information to help each particle moving in search space towards the global optimal solution. The algorithm is initialized with a population of random particles within the feasible region, and then each particle uses experience of its own best location and the social best position to adjust its trajectory to search for a globally optimal solution in each evolutionary step \cite{Ebbe12, Kenn95}. Denote $\pmb{X} = \left\{ {{\pmb{F}_k}} \right\}_{k = 1}^K$ as a set of precoding matrices. At the begin of the algorithm, the $S$ particles are randomly initialized. The position and velocity of the $i$-th particle can be represented by $\pmb{X}_i$ and $\pmb{V}_i$, respectively. The fitness of each partible can be calculated according to the cost function of the optimization problem. For each particle of the warm, the best previous moved position of the particle $i$ is denoted as its best individual position $\pmb{X}_{pbest,i}$. The best position of the entire swarm can be represented by $\pmb{X}_{gbest}$. Based on the cognitive and social information, at the $\kappa$-th iteration, the velocity and position of the particle $i$ is adapted according to two following equations \cite{Ebbe12, Kenn95}:
\begin{equation}\label{PSO:Velocity}
\begin{array}{ll}
  \pmb{V}_i^{(\kappa+1)} =& c_0 \pmb{V}_i^{(\kappa)} + c_1r_{1,i}^{(\kappa)}({\pmb{X}_{pbest,i}^{(\kappa)}} - \pmb{X}_i^{(\kappa)}) \\
  & + c_2r_{2,i}^{(\kappa)}(\pmb{X}_{gbest}^{(\kappa)} - \pmb{X}_i^{(\kappa)}),
\end{array}
\end{equation}
\begin{equation}\label{PSO:Position}
\pmb{X}_i^{(\kappa+1)}=\pmb{X}_i^{(\kappa)}+ \pmb{V}_i^{(\kappa+1)},
\end{equation}
where $c_0$ is called the inertial weight that controls the effect of the previous velocity of a particle on its current one. $c_1$ and $c_2$ are known as the cognitive and social parameters, respectively, which control the maximum step size. The random numbers $r_{1,i}$ and $r_{2,i}$  are independently uniformly distributed in the interval $[0,1]$. It is critical to choose these parameters since they affect the convergence characteristic of the PSO algorithm. An early convergence may occur with too small values of $c_0$ while a slow convergence may be result from $c_0$ being too large. In this paper, we select the values of $c_0=0.7$, $c_1=1.494$ and $c_2=1.494$ according to experimental tests in \cite{Robi04} where these chosen parameters can commonly offer the good convergence characteristic.

Although Eq. \eqref{PSO:Velocity} and \eqref{PSO:Position} perform a global search for unconstrained optimization problem, the design problem \eqref{FkOpt} is the constrained optimization one. Therefore, it is essential to develop a new constrained PSO algorithm. To deal with constraints, there are typically two approaches, namely the projection and penalty methods \cite{Boyd04}. The basic idea of projection method is to project the solution of the associated unconstrained optimization problem into the feasible region of the constraints while the penalty method focuses on transforming the constrained optimization problem into unconstrained optimization one by using the penalty function. Since choosing appropriate penalty parameters is difficult, this paper integrates the projection methods into the standard unconstrained PSO algorithm for finding the optimal precoders. We define the feasible region $\mathcal{F}$ of the optimization problem \eqref{FkOpt} as
\begin{align}
{\mathcal{F}}=\left\{ \pmb{X}=\{\pmb{F}_k\}_{k=1}^K: \sum\limits_{k=1}^K\tr(\pmb{F}_k\pmb{F}_k^H) \leq P_{max} \right\}.
\end{align}
If particle $\pmb{X}_i$  is not in $\mathcal{F}$, then we project it back into the feasible region $\mathcal{F}$. The projection of $\pmb{X}_i$ into $\mathcal{F}$ is defined as
\begin{equation}
\hat{\pmb{X}}_i=\Pi(\pmb{X}_i)
\end{equation}
which is mathematically expressed as
\begin{equation}\label{proj}
\Pi(\{\pmb{F}_{k}\}_{k=1}^K)=\text{arg}\min_{\{\hat{\pmb{F}}_{k}\}_{k=1}^K \in \mathcal{F}} \sum\limits_{k=1}^K||\hat{\pmb{F}}_{k}-\pmb{F}_{k}||_F
\end{equation}
where $\hat{\pmb{X}}_i=\{\hat{\pmb{F}}_{k}\}_{k=1}^K$ is the closed point in $\mathcal{F}$ to $\pmb{X}_i$. Eq. \eqref{proj} is equivalently rewritten as
\begin{subequations}\label{projtr}
\begin{eqnarray}
&  \underset{\{\hat{\pmb{F}}_{k}\}_{k=1}^K}{\min} & \sum\limits_{k=1}^K\tr((\hat{\pmb{F}}_{k}-\pmb{F}_{k})(\hat{\pmb{F}}_{k}-\pmb{F}_{k})^H) \\
& \text{s.t.} & \sum\limits_{k=1}^K\tr(\hat{\pmb{F}}_{k}\hat{\pmb{F}}_{k}^H) \leq P_{max}.
\end{eqnarray}
\end{subequations}
It is obvious that problem \eqref{projtr} is convex optimization. The Lagrangian expression of problem  \eqref{projtr} is defined as
\begin{align}
{\mathcal{L}}(\{\hat{\pmb{F}}_{k}\}_{k=1}^K, \mu)= &\sum\limits_{k=1}^K\tr((\hat{\pmb{F}}_{k}-\pmb{F}_{k})(\hat{\pmb{F}}_{k}-\pmb{F}_{k})^H) \\ \nonumber &+\mu(\sum\limits_{k=1}^K\tr(\hat{\pmb{F}}_{k}\hat{\pmb{F}}_{k}^H)-P_{max})
\end{align}
where $\mu$ is the Lagrangian dual variables associated with the power constraint. The optimal solutions of problem \eqref{projtr} must satisfy the set of Karush-Kuhn-Tucker (KKT) conditions
\begin{subequations}\label{KKT}
\begin{eqnarray}
 (1+\mu)\hat{\pmb{F}}_{k}-\pmb{F}_{k}&=&0, \quad k=1,2,...,K \\
 \sum\limits_{k=1}^K\tr(\hat{\pmb{F}}_k\hat{\pmb{F}}_k^H)&=&P_{max}\\
 \mu &\geq& 0
\end{eqnarray}
\end{subequations}
which immediately implies that
\begin{equation}\label{proresult}
\hat{\pmb{F}}_{k}=\sqrt{\frac{P_{max}}{\sum\limits_{k=1}^K\tr(\pmb{F}_k\pmb{F}_k^H)}}\pmb{F}_{k}, \quad k=1,2,...,K.
\end{equation}
Consequently, the constrained PSO using the projection method for finding the optimal transceivers can be described in Algorithm \ref{Alg:PSO} where $K_{\max}$ is the maximum number of iterations.
\begin{algorithm}[shadow]
\caption{: PSO for Downlink MIMO Broadcast Channels}\label{Alg:PSO}
\begin{algorithmic}[1]
\STATE Declare variables $\pmb{X} = \left\{ {{\pmb{F}_k}} \right\}_{k = 1}^K$.
\STATE Inputs: $d,\left\{ {{\pmb{H}_k}} \right\}_{k = 1}^K,{\sigma _k},{N_{{t}}},{N_{{r}}},{P_{\max }}, K_{\max}, S, c_0, c_1, c_2$, $\left\{ {\omega_k} \right\}_{k = 1}^K$, $\pmb{X}_{BD}$, where $\pmb{X}_{BD}$ is the solution of BD method.
\STATE Set the iteration index $\kappa=0$.
\STATE Generate a swarm of random solutions satisfying the power constraints $\left\{ {{\pmb{X}_i}} \right\}_{i = 2}^S$, where $S$ is the swarm size. Set $ {{\pmb{X}_1}}  = \pmb{X}_{BD}$.
\STATE Initialize for each particle ${\pmb{X}_{pbest,i}} = \left\{ {{\pmb{X}_i}} \right\}$ and ${\pmb{V}_i} = 0$.
\STATE Calculate $\pmb{W}_k$ from Eq. \eqref{Wk:Wiener} for $k=1,2,..K$.
\STATE Evaluate the cost function of each particle $\left\{ {{\mathcal{R}}\left( {{\pmb{X}_i}} \right)} \right\}_{i = 1}^S$
\WHILE{$\kappa < K_{\max }$}
\STATE Select the leader $\pmb{X}_{gbest}$.
\STATE For each particle, update the velocity and it position by Eqs. \eqref{PSO:Velocity} and  \eqref{PSO:Position}.
\STATE If $\pmb{X}_i$ is not into the feasible region, then use  Eq. \eqref{proresult} to  project $\pmb{X}_i$ into the feasible region:
 ${\pmb{X}_i} \leftarrow \Pi(\pmb{X}_i)$.
\STATE Calculate $\pmb{W}_k$ from Eq. \eqref{Wk:Wiener} for $k=1,2,..K$.
\STATE Evaluate the cost function of each particle from Eq. \eqref{objFkOpt}.
\STATE Update the $\pmb{X}_{pbest,i}$ for each particle.
\STATE Update $\kappa=\kappa+1$ then move to the next "Do" task
\ENDWHILE
\end{algorithmic}
\end{algorithm}

It is worth noting from Algorithm \ref{Alg:PSO} that the particle with the highest WSR is chosen at each iteration and, thus, the objective \eqref{objFkOpt} is non-decreasing over iteration, i.e.,
\begin{equation}
\mathcal{R}(\pmb{X}_{gbest}^{(\kappa+1)}) \geq \mathcal{R}(\pmb{X}_{gbest}^{(\kappa)}).
\end{equation}
However, initial particle positions affect the convergence speed of the PSO algorithm. In this paper, we propose the use of the solution of the BD method as an initial particle while other initial particles are randomly chosen. It means that our algorithm aims at searching the precoders which offers the higher sum-rate than those of BD methods since the precoders designed by the BD method are suboptimal solutions \cite{Shen06}. It should be emphasized that a key advantage of the proposed PSO algorithm over the BS schemes is that the PSO algorthim does not impose any restriction of the number of users and antennas. The speed of convergence of Algorithm \ref{Alg:PSO} shall be illustrated in the simulation results.
The computational complexity of Algorithm \ref{Alg:PSO} mainly relies on the computation of the initial particle from the BD scheme and the evaluation of the objective function for a swarm of $S$ particles. For the BD method, the computational complexity which consists of the SVD computation and waterfilling algorithm is ${\mathcal{O}}(KN_t^3+K^2d^2)$ \cite{Sung09}. The major computation complexity of the PSO algorithm consists of computing matrix multiplications, matrix inversion, and determinant in Eqs. \eqref{Cap:MSk} and \eqref{Wk:Wiener}. Note that computational complexity of computing the inversion and determinant of $N \times N $ matrix is ${\mathcal{O}}(N_t^3)$. Thus, evaluating the objective function approximately yields the complexity of ${\mathcal{O}}(K^2N_t^3)$. With $S$ particles and the maximum number of iterations $K_{\max}$, the major computational complexity of Algorithm \ref{Alg:PSO} is ${\mathcal{O}}(S K_{\max} K^2N_t^3)$.
\section{Simulation Results}
\label{sec:Results}
In this section, numerical simulation results are provided to evaluate the performance of our proposed PSO algorithm for downlink multiuser MIMO interference channels. The performance of the proposed method is compared to the BD method \cite{Spen04} in terms of the total WSR. We denote the $K$ user MIMO downlink channels with $N_t$ transmit antennas at the BTS and $N_r$ receive antennas at each MS by $\left( N_{t} \times N_{r}, d \right)^K$ \cite{Papailiopoulos2012}. In simulations, noise variances are normalized $\sigma_k^2=\sigma^2=1$. The Rayleigh fading channel coefficients are generated from the complex Gaussian distribution ${\mathcal{CN}}(0,1)$. Thus, we define $\text{SNR}=\frac{P_{\max}}{\sigma^2}$. All the numerical results are averaged over the $200$ channel realizations. In the following experiments, we select empirically the number of swarm size $S=700$, hence, the local global optimum can be guaranteed and if $S$ is chosen larger or $S \rightarrow \infty$, we can obtain the global solution.

\begin{figure}[htb!]
\begin{center}
\epsfxsize=9cm
\leavevmode\epsfbox{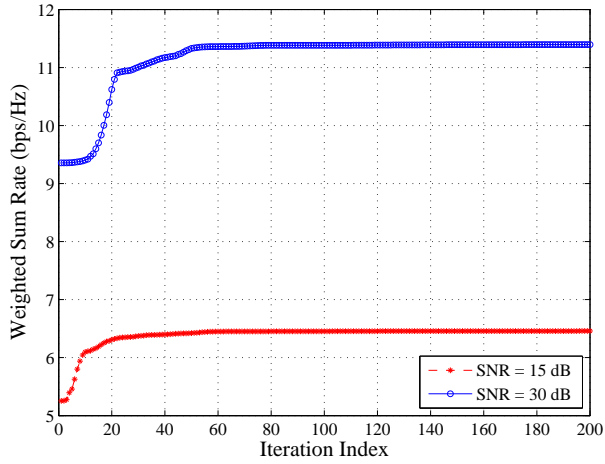}
\caption{The convergence behavior of the proposed algorithm.}
\label{fig:PSOconverge}
\end{center}
\end{figure}

First, we investigate the convergence characteristic of the proposed PSO algorithm.  We run the simulation for a ${\left( 6 \times 2,1 \right)^3}$ system for a random channel realization. Weighed factors for $3$ users are $[\omega_1, \omega_2, \omega_3] = [0.1, 0.2, 0.7]$. The evolution of the objective function (weighted sum-rate) over iterations is illustrated in Fig. \ref{fig:PSOconverge}. As can been seen from Fig. \ref{fig:PSOconverge}, the total WSR is significantly improved after several first iterations. The algorithm is converged to a fixed point in less than $150$ iterations.

\begin{figure}[!t]
\begin{center}
\epsfxsize=9cm
\leavevmode\epsfbox{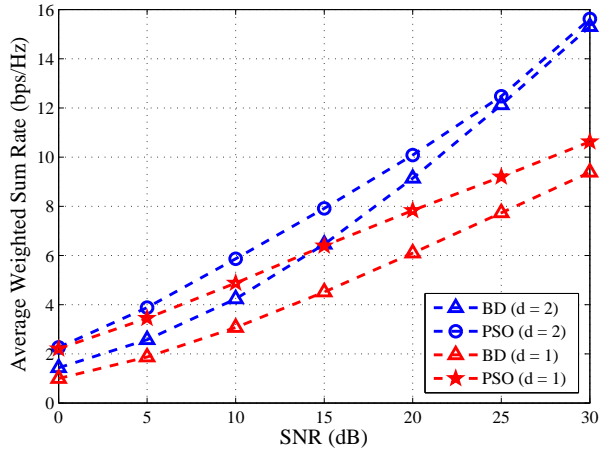}
\caption{The average weighted sum-rate versus SNR for ${\left( 6 \times 2,\{1,2\} \right)^3}$ systems.}
\label{fig:PSO62}
\end{center}
\end{figure}

Now, we compare the WSR of our proposed algorithm with that of the BD method. First, we consider a ${\left( 6 \times 2,\{1,2\} \right)^3}$ system where the condition $N_t-(K-1)N_r \geq d$ is satisfied. Thus, the BD scheme can completely cancel inter-user interference.  The WSR is plotted in Fig. \ref{fig:PSO62}. It has been revealed from Fig. \ref{fig:PSO62} that the proposed algorithm offers a WSR  performance improvement as compared to the BD method. The sum rate improvement is significant at low SNR while the performance gap is smaller at high SNR. The reason is that at high SNR the system is interference-limited and the BD scheme approaches the optimal solution since it can cancel all interference.

 We also evaluate the WSR performance for different systems of ${\left( 4 \times 2,\{1,2\} \right)^3}$. The average WSR is shown in Fig. \ref{fig:PSO42}. It can be observed that the WSR performance of the BD method is not increased with an increasing SNR at the interference-limited region since the BD method cannot completely cancel all interference for these systems with $N_t-(K-1)N_r \leq d$. In contrast, our proposed algorithm directly maximizes the WSR and does not impose the restriction on the number of antennas and MSs. It can been seen from Fig. \ref{fig:PSO42} that our method consistently outperforms the BD method, especially for high SNR for both cases of $d=1$ and $d=2$.
 \begin{figure}[htb!]
\begin{center}
\epsfxsize=9cm
\leavevmode\epsfbox{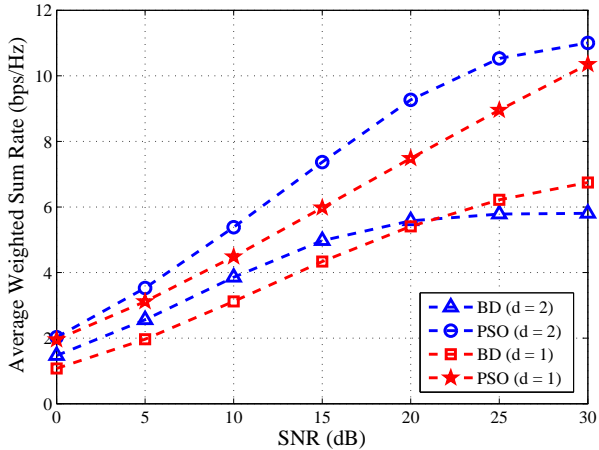}
\caption{The average weighted sum-rate versus SNR for ${\left( 4 \times 2,\{1,2\} \right)^3}$ systems.}
\label{fig:PSO42}
\end{center}
\end{figure}
\section{Conclusion}
\label{sec:Conclusion}
In this paper, we have studied the joint optimal transceiver designs for downlink multiuser MIMO interference channels. We have formulated the problem of finding the precoding and decoding matrices to achieve the maximization of the weighted sum rate subject to the total power constraints imposed on the BTS. The proposed approach is to employ an efficient stochastic PSO algorithm to obtain the optimal solutions. The numerical results indicate that the proposed PSO algorithm outperforms the BD method in terms of weighted sum rate. The weighted sum rate performance improvement is highly significant when the number of users or receive antennas increases and the BD method cannot completely cancel interference. In addition, the numerical experiments have shown that the proposed methods can converge to a fixed point in less than a hundred and fifty iterations while the computational complexity per iteration is relatively low and affordable in practical applications.
\section*{Acknowledgement}
This research is funded by Vietnam National Foundation for Science and Technology Development (NAFOSTED) under grant number 102.04-2013.46.
%

%
%

\end{document}